\documentclass[]{spie}  

\usepackage{amsmath,amsfonts,amssymb}
\usepackage{graphicx}
\usepackage[colorlinks=true, allcolors=blue]{hyperref}
\usepackage{float} %

\usepackage{array}    
\usepackage{booktabs} 
\usepackage{tabularx} 
\usepackage[table]{xcolor}
\usepackage{subcaption}

\title{NCF: Neural Correspondence Field for\\ Medical Image Registration}

\author {Lei Zhou}
\author{Nimu Yuan}
\author{Katjana Ehrlich}
\author {Jinyi Qi}
\affil{Department of Biomedical Engineering, University of California, Davis, CA, USA}

\authorinfo{Further author information: (Send correspondence to Jinyi Qi, E-mail: qi@ucdavis.edu)}

\begin{document} 
\maketitle

\begin{abstract}
Deformable image registration is a fundamental task in medical image processing. Traditional optimization-based methods often struggle with accuracy in dealing with complex deformation. Recently, learning-based methods have achieved good performance on public datasets, but the scarcity of medical image data makes it challenging to build a generalizable model to handle diverse real-world scenarios. To address this, we propose a training-data-free learning-based method, Neural Correspondence Field (NCF), which can learn from just one data pair. Our approach employs a compact neural network to model the correspondence field and optimize model parameters for each individual image pair. Consequently, each pair has a unique set of network weights. Notably, our model is highly efficient, utilizing only 0.06 million parameters.
Evaluation results showed that the proposed method achieved superior performance on a public Lung CT dataset and outperformed a traditional method on a head and neck dataset, demonstrating both its effectiveness and efficiency.
\end{abstract}


\section{INTRODUCTION}
\label{sec:intro} 

Image registration plays a critical role in various medical image applications, including diagnosis, treatment planning, and treatment monitoring. It involves aligning multiple images into a shared coordinate system, which is essential for comparing anatomical structures and tracking changes over time. 
Typically, registration can be divided into three main categories: rigid, affine, and deformable registration. Rigid registration aligns images using translations and rotations, while affine registration includes scaling and shearing. Deformable registration captures complex, non-rigid transformations, crucial for modeling anatomical motion or deformation between images.

Traditional deformable image registration (DIR) methods, such as Elastix\cite{klein2009elastix}, rely on iterative optimization but often struggle with complex deformations\cite{tian2023same++}. In recent years, deep learning-based methods\cite{balakrishnan2019voxelmorph,chen2021vit} have achieved remarkable results in DIR. However, these methods\cite{zheng2024residual} rely heavily on large datasets to improve generalization ability. Unlike other fields where data are abundant and easily accessible, medical image data is typically limited in scale. As a result, deep learning-based algorithms are generally restricted to registering images that resemble the training data. For example, a model trained on CT images of one body region may not generalize well to another.

\begin{figure}[t] 
    \centering
    \begin{subfigure}[b]{0.45\textwidth}
        \centering
        \includegraphics[width=\textwidth]{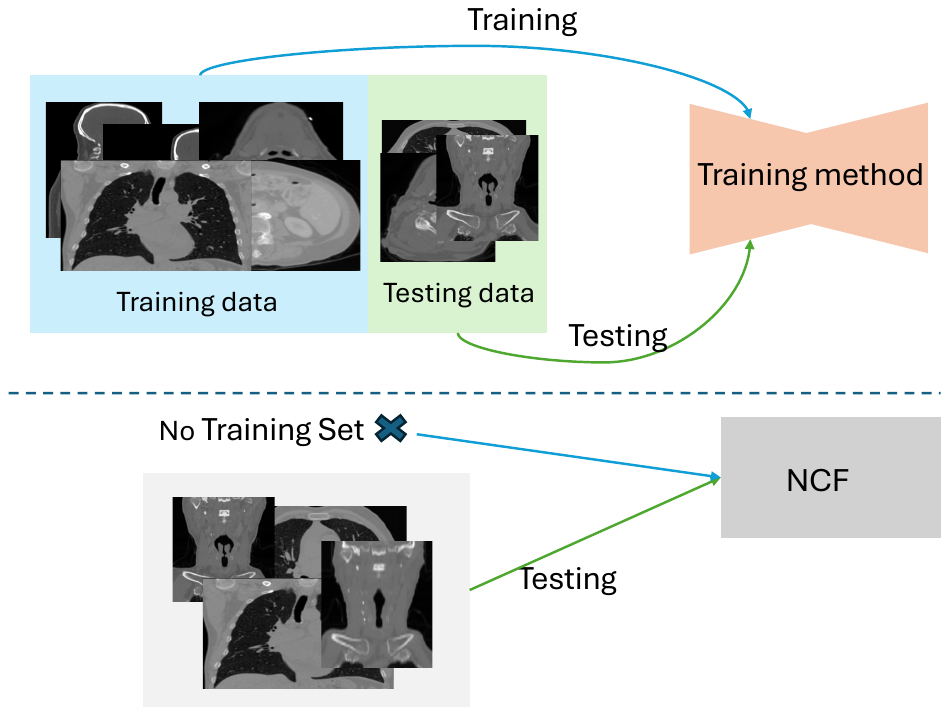}
        \caption{}
        \label{fig:training_free}
    \end{subfigure}
    \hfill
    \begin{subfigure}[b]{0.45\textwidth}
        \centering
        \includegraphics[width=\textwidth]{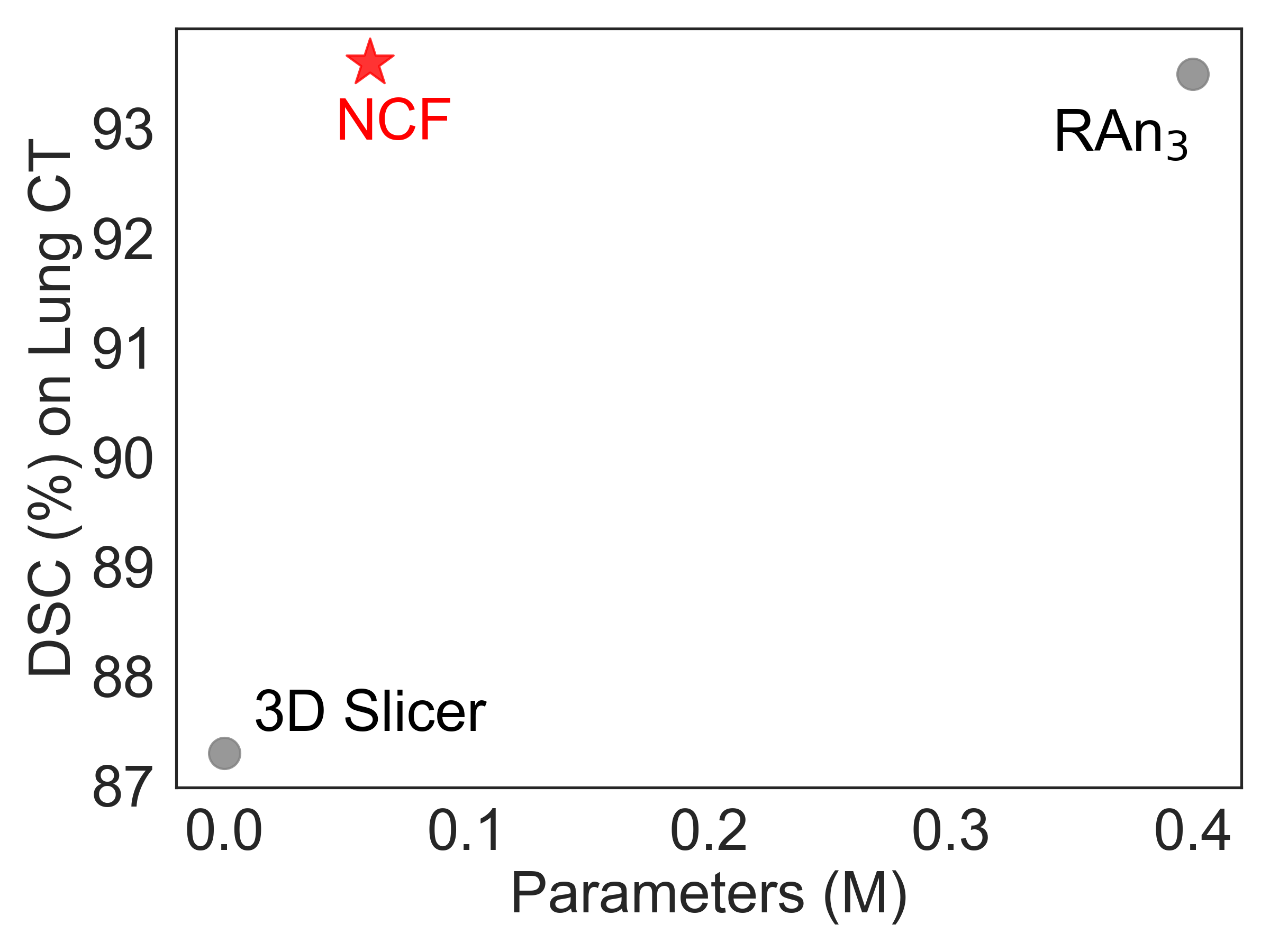}
        \caption{}
        \label{fig:parameters}
    \end{subfigure}
    \caption{Features and advantages of the proposed method. (a) Comparison with learning-based methods in terms of the need for a training dataset. (b) Comparison of the registration performance (measured by Dice Similarity Coefficient (DCS) of segmentation masks) vs. the number of network parameters.}
    \label{fig:overall}
\end{figure}

In this paper, we propose a novel training-data-free network to learn the deformable correspondence field for CT images. Unlike deep learning-based methods that require extensive datasets for training and validation, our method optimizes directly on an individual image pair, as shown in Fig. \ref{fig:training_free}, eliminating the need for additional data.
Our method is highly data-efficient, requiring only one image pair to train the network. Furthermore, the network parameters are specific to each training sample. Thus, the generalization ability of NCF stems from the model architecture itself rather than the pre-trained network parameters. To support this framework, we design a super lightweight neural network along with an efficient training strategy. Experimental results showed that our model outperformed 3D Slicer, a widely used traditional DIR tool, and performed comparably to deep learning-based methods that rely on extensive training data (Fig. \ref{fig:parameters}).

\section{Related Work}

Deformable image registration has been a critical task in medical image analysis. A wide range of medical image registration methods has been developed, broadly categorized into traditional optimization-based techniques and more recent deep learning-based approaches.  

\noindent
{\bf Optimization-based medical image registration.}
Early medical image registration employed iterative optimization techniques to handle the deformable alignment of medical images. Elastix \cite{klein2009elastix}, an open-source registration tool, provides an intensity-based registration framework, facilitating the registration of images across different modalities and time points. Modat et al.\cite{modat2010fast} proposed a GPU-based approach for fast free-form deformation (FFD), leveraging parallel computing to enhance efficiency. SyN\cite{avants2008symmetric} was developed to maximize cross-correlation within the framework of topology-preserving registration, providing a variational optimization framework grounded in the Euler-Lagrange equations. However, these methods are often computationally expensive and struggle with complex deformable image registration scenarios.

\noindent
{\bf Deep learning-based medical image registration.}
Deep learning-based image registration introduces data-driven approaches to complex mappings directly from data. Researchers have explored supervised or weakly-supervised methods to learn deformation fields \cite{yang2017fast, gao2024mairnet, xu2019deepatlas}. However, obtaining pixel-wise labels and segmentation information is often costly and labor-intensive. Recent advancements in deep learning-based registration methods aim to eliminate the dependence on labels by adopting unsupervised training approaches \cite{meng2024correlation, zheng2024residual, chen2022transmorph}. VoxelMorph \cite{balakrishnan2019voxelmorph} employed an encoder-decoder CNN to predict non-linear local deformations. TransMorph \cite{chen2022transmorph} integrated Swin Transformers and ConvNets to model long-range dependencies. RAN \cite{zheng2024residual} targeted complex motion patterns and designed the Residual Aligner module to refine predicted motions. CorrMLP \cite{meng2024correlation} leveraged multi-window MLPs to capture multi-range dependencies. These deep learning-based methods have significantly advanced image registration. However, they typically require large amounts of training data for effective generalization. For instance, CorrMLP utilized 2,656 brain MRI images for training. Furthermore, the generalization of deep learning-based methods is constrained to specific anatomical structures present in the training data.

To address these challenges, our approach combines deep learning with iterative optimization, reducing the dependency on large training datasets while maintaining the state-of-the-art performance with significantly fewer parameters.

\section{Method}

\subsection{Network Design}

\begin{figure}[t]
    \centering
    \includegraphics[width=0.8\textwidth]{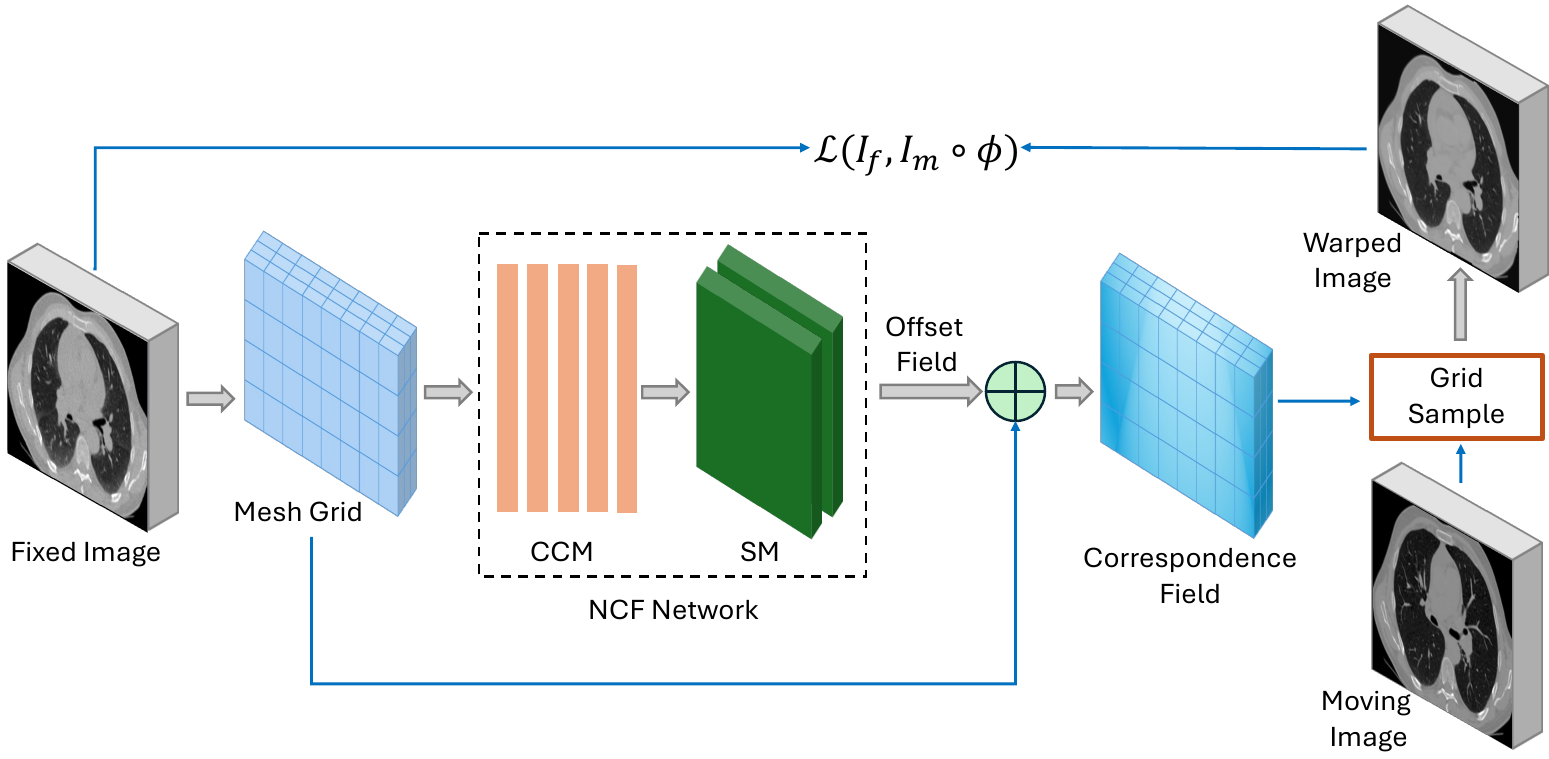} 
    \caption{{\bf Architecture}. The proposed method, NCF, consists of two parts: the CCM, represented in orange, is a 5-layer MLP network, and the SM, represented in green, is a 2-layer 3D CNN network. The loss function $\mathcal{L}$ evaluates the difference between the warped image $I_m \circ \Phi$ and the fixed image $I_f$.}
    \label{fig:architecture}
\end{figure}

The architecture of the proposed Neural Correspondence Field (NCF) is illustrated in Fig. \ref{fig:architecture}. 
It consists of two primary modules: a Coarse Correspondence Module (CCM) and a Smooth Module (SM). The CCM consists of a 5-layer multi-layer perceptron (MLP) and is fed with the coordinates of each voxel. Given a pair of fixed and moving 3D images, the CCM is designed to learn a coarse deformation field for each voxel in the moving image. By using a globally-shared MLP, the CCM tends to produce a locally smooth output. 
To further enhance the smoothness of the deformation field, the SM, consisting of a 2-layer 3D convolutional neural network (CNN), is appended to post-process the coarse deformation field. 
CNNs, as demonstrated by the Deep Image Prior (DIP)\cite{ulyanov2018deep} concept, possess inherent denoising and smoothing capabilities due to their intrinsic architecture and the training process. The SM leverages these capabilities to refine and smooth the learned offset field, ensuring a more accurate and reliable result.

\begin{figure}[t]
    \centering
    \includegraphics[width=0.8\textwidth]{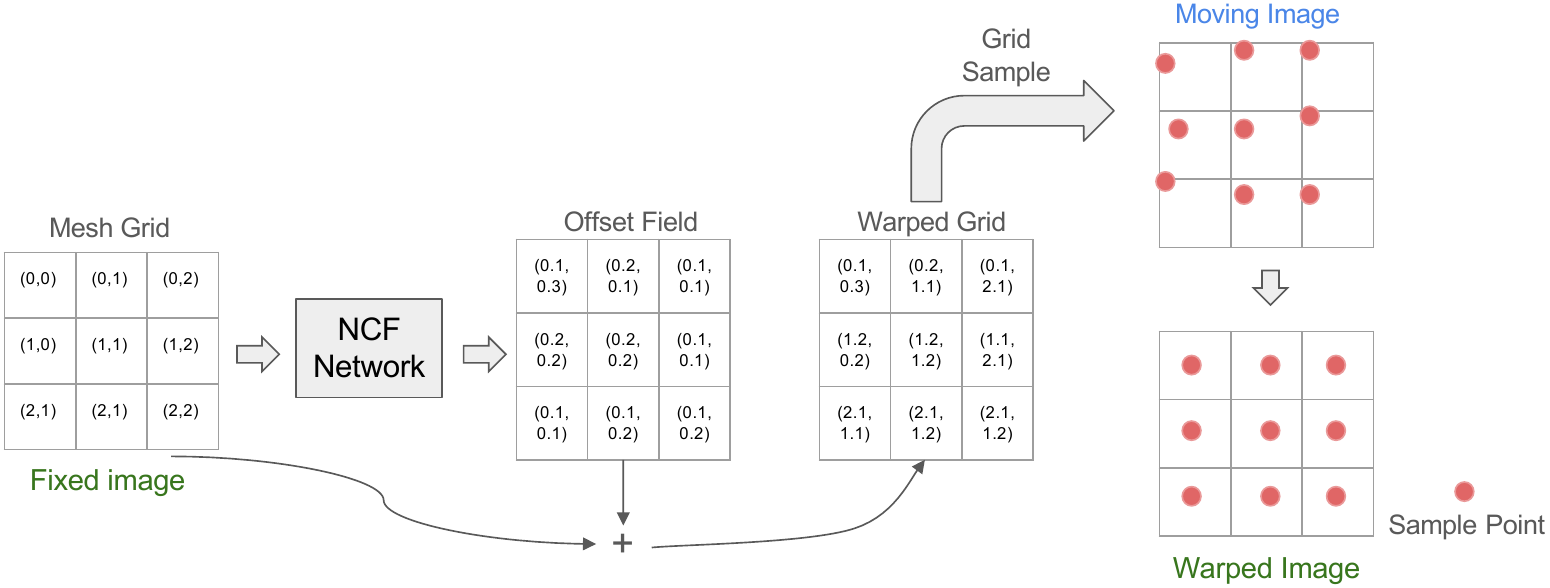} 
    \caption{A 2D illustration of generating a warped image through grid sampling of the moving image based on the warped grid.}
    \label{fig:example_ncp}
\end{figure}

 The input of our NCF network is a 3D mesh grid of dimensions $(3,W,H,D)$, where $W,H,D$ represent the width, height, and depth of the fixed image $I_{f}$, respectively, and the output is an offset field of the same size as the mesh grid. By adding the mesh grid and the offset field together, a correspondence field is obtained. By sampling the moving image based on the correspondence field, a warped image, denoted as $I_{m} \circ \phi$, can be generated. This approach enables the network to be trained in a self-supervised manner.
A 2D illustration of the grid sampling process is shown in Fig. \ref{fig:example_ncp}.

\subsection{Loss Design}
The total loss functions comprise three parts: photometric loss, Structural Similarity Index Measure (SSIM) loss, and occupancy loss. Photometric loss, $\mathcal{L}_{photometic}$, computes the mean squared difference between a fixed image and a warped image. 
The SSIM loss, $\mathcal{L}_{SSIM}$, incorporates luminance, contrast, and structural information to provide a more holistic assessment of image quality. 
The occupancy loss, $\mathcal{L}_{occupancy}$, is employed to assess the smoothness of an occupancy tensor $B$, which represents the frequency of each voxel in the moving image being visited.
It is quantified by calculating the Root Mean Square Deviation (RMSD) of the differences in occupancy values along the $x$, $y$, and $z$ dimensions.
The final loss function $\mathcal{L}(I_f, I_m \mathbin{o} \Phi)$ is a sum of the three loss functions:
\begin{equation}
    \mathcal{L}(I_f, I_m \mathbin{o} \Phi) = \alpha \mathcal{L}_{photometic} + \beta \mathcal{L}_{SSIM} + \gamma \mathcal{L}_{occupancy}.
\end{equation}

\noindent
We empirically set $\alpha=1$, $\beta=1$, and $\gamma=0.1$ in our experiment.

\section{Experiments}
\label{sec:Implementation}

\subsection{Implementation Details}
Our model was implemented using PyTorch, and trained on an A100 GPU with 80GB memory. Training employed the Adam optimizer scheduled by CosineAnnealingLR with an initial learning rate of 1e-3 and a final value of 1e-6. A batch size of 1 was used throughout the training process.

\subsection{Dataset}

\subsubsection{Unpaired/paired Lung CT dataset}
The Lung CT dataset is provided by the Learn2Reg Challenge \cite{hering2020learn2reg}. Each image has a spatial resolution of $1.75 \times 1.25 \times 1.75$  $mm^3$
  and a size of $192 \times 192 \times 208$. We compared our NCF method with two existing methods: a traditional method, 3D Slicer \cite{fedorov20123d} with the Elastix \cite{klein2009elastix} library, and a learning-based method, specifically RAN \cite{zheng2024residual}. We report the Dice Similarity Coefficient (DSC) of the segmented lung masks and the number of network parameters for each method to evaluate and compare their performance.

Notably, RAN used landmarks in intra-patient experiments during training, whereas 3D Slicer and our NCF method did not. 
\subsubsection{Private Head and Neck data}
Our private dataset comprises conventional CT and PET/CT scans of 14 head and neck cancer patients. 
The images were acquired at different pre-surgery time points, leading to variations in patient posture, potential structural changes, and differences in image size and spatial resolution.

This dataset provides a robust testbed for evaluating the performance of our NCF method under real-world conditions.

\subsection{Results}

\subsubsection{Result on the Lung CT dataset}
We conducted experiments on intra-patient 
and inter-patient 

scenarios, following the default settings outlined in the RAN\cite{zheng2024residual} method and Learn2Reg Challenge. For the intra-patient experiments, we selected the first 8 samples for individual learning and testing. 

For inter-patient conditions, we used 56 pairs for evaluation. Table \ref{tab:lungCTComparison_inter_intra} shows the results for both intra-patient and inter-patient registrations. In both cases, our method achieved the best performance with the fewest parameters. It is worth noting that our method did not use any information from the training set. For the intra-patient study, our method outperformed RAn$_{3}$, which used landmarks and training data to improve its model, setting a new state-of-the-art in the Lung CT paired registration task.

\subsubsection{Result on Head and Neck dataset}

\begin{figure}[t]
    \centering
    \includegraphics[width=0.5\textwidth]{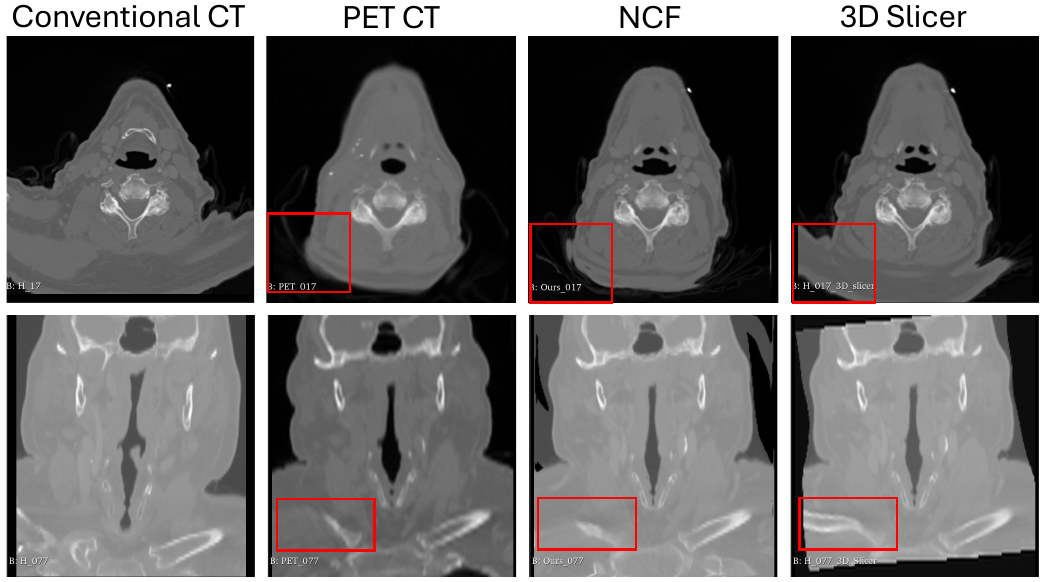} 
    \caption{Comparing the results of our NCF method and the 3D Slicer on the head and neck dataset. Red rectangles mark the areas where our method outperformed the 3D Slicer.}
    \label{fig:Private_Data}
\end{figure}

For the head and neck dataset, the registration was performed from conventional CT to PET/CT. Due to the lack of labels in the dataset, we provide visual inspection rather than quantitative metrics in Fig. \ref{fig:Private_Data}. Visually, the results of our method align better with the target PET/CT image than the 3D Slicer output. 

\begin{table}[h]
    \centering
    \caption{Comparison between our NCF method and others on the Lung CT dataset. We report the Dice Similarity Coefficient (DSC$\uparrow$) of the segmented lung masks and the number of parameters. Our method achieves the best results with the fewest learnable parameters.}
    \begin{tabular}{>{\centering\arraybackslash}p{2cm} >{\centering\arraybackslash}p{6cm}  >{\centering\arraybackslash}p{4cm} > {\centering\arraybackslash}p{3cm}}
        \toprule
        \textbf{Model}  & \textbf{Inter- and Intra-patient (DSC)} & \textbf{Intra-patient (DSC)} & \textbf{Parameter (M)} \\
        \midrule
        3D Slicer\cite{klein2009elastix} & 0.873 & 0.883 & - \\
        RAn$_{3}$\cite{zheng2024residual} &  0.935 & 0.935 & 0.4 \\
        \rowcolor{gray!25} \textbf{NCF} &  \textbf{0.936} & \textbf{0.942} & \textbf{0.06} \\
        \bottomrule
    \end{tabular}
    \label{tab:lungCTComparison_inter_intra}
\end{table}

\section{Conclusion}
In this paper, we propose a novel, training-data-free learning-based network, called NCF, for deformable image registration (DIR). Our lightweight model eliminates reliance on extensive training datasets. Experimental studies using public and private datasets demonstrated that the proposed NCF possessed excellent generalization ability and achieved superior performance compared to a pre-trained deep network model and a traditional optimization-based method using a few parameters.

\acknowledgments 

This work is supported by NIH under grant P41 EB032840 and R01 CA187427.

\bibliography{report} 
\bibliographystyle{spiebib} 

\end{document}